\documentstyle[12pt]{article}
\begin{document}
\def\baselinestretch{1.5}
\newcommand{\lb}{\label}
\newcommand{\be}{\begin{equation}}
\newcommand{\ee}{\end{equation}}
\newcommand{\fr}{\frac}
\newcommand{\D}{\delta}
\newcommand{\sig}{\sigma}
\newcommand{\del}{\partial}
\newcommand{\wv}{\wedge}
\newcommand{\al}{\alpha}
\newcommand{\la}{\lambda}
\newcommand{\La}{\Lambda}
\newcommand{\ep}{\epsilon}
\newcommand{\pr}{\prime}
\newcommand{\ti}{\tilde}
\newcommand{\om}{\omega}
\newcommand{\OO}{\Omega_{BRST}}
\newcommand{\bi}{\bibitem}

\title{
\large
SOLUTION OF THE MASTER EQUATION IN
TERMS OF THE ODD TIME FORMULATION}

\author{
\normalsize
\"{O}mer F. DAYI \\
\normalsize \it
University of Istanbul,
Faculty of Science,  \\
\normalsize \it
Department of Physics,
Vezneciler, 34459 Istanbul,
Turkey\thanks{Address after 1 August 1992:
ICTP, P.O.Box 586, 34100-Trieste,Italy;
bitnet: dayi@itsictp} }
\date{}
\maketitle
\begin{abstract}
A systematic way of formulating the Batalin-Vilkovisky method of
quantization was obtained in terms of the ``odd time'' formulation.
We show that in a class of gauge theories it is possible to
find an ``odd time lagrangian'' yielding, by a Legendre transformation,
an ``odd time hamiltonian'' which is the minimal solution of  the
master equation. This constitutes a very simple method of finding
the minimal solution of the
master equation which is usually a tedious task.
To clarify the general
procedure we discussed its application
to Yang-Mills theory, massive
(abelian) theory in Stueckelberg formalism, relativistic particle
and the self-interacting antisymmetric tensor field.
\end{abstract}
\pagebreak

The formulation of the Batalin-Vilkovisky (BV) method of
quantization\cite{BV} in terms of a Grassmann odd parameter
behaving as time (``odd time") was helpful to
derive systematically the ``ad hoc" definitions of
Batalin and Vilkovisky\cite{o1}. In Ref.\cite{o1}
existence of an
appropriate lagrangian for the odd time formulation
(``odd time lagrangian") was assumed. Although it is possible
to write an odd time lagrangian by using the hamiltonian formalism
of the BV method\cite{hbv}, it does not give new
hints about finding a solution of the master equation
of the lagrangian formalism.
In the hamiltonian formalism a general solution of
the master equation is available in terms of the
Becchi-Rouet-Stora-Tyutin (BRST)-charge
which gives a vanishing generalized Poisson bracket with
itself.

Recently, the extended form method developed in Ref.\cite{cc}
was utilized to write actions  for topological quantum field
theories which lead to solutions of the master
equation\cite{ikemori}. In fact,
those actions are nothing but odd time lagrangians of the related theories.
Inspired by these, we showed that for a class of gauge theories
it is possible to write an odd time lagrangian which directly
leads to the minimal solution of the master equation in terms
of ``odd canonical formulation''.

First we recall the basic concepts of the odd time formulation
of the BV method and show that odd canonical formalism
is similar to the normal one. After giving the rules of finding
odd time lagrangian beginning from the initial gauge theory
action, we discuss the conditions which they should satisfy in order
to yield a solution of the master equation. We illustrate the
method by its application
to Yang-Mills theory, massive
(abelian) theory in Stueckelberg formalism, relativistic particle
and the self-interacting antisymmetric tensor field.

When we deal with a gauge theory we can introduce
the  odd time $\tau_0$
(a parameter possessing odd Grassmann parity),
such that the change of a function f by
the BRST-charge $\OO $,
is written symbolically as
\[
\OO f=\fr{\del f(\tau_0)}{\del \tau_0}.
\]
We assume that there exists an odd time
lagrangian
$ L( \phi(\tau_0), \dot{\phi}(\tau_0) ) $,
which carries  information about
the BRST transformations. $\phi$ includes the original
fields of the started gauge theory, and the related
ghost fields;
      $ \dot{\phi}(\tau_0) \equiv \del \phi (\tau_0) / \del \tau_0$.
The ``odd time canonical momentum" which results from this lagrangian is
\be
\Pi (\tau_0)=\fr{\del L (\phi (\tau_0) ,\dot{\phi}(\tau_0))}
{\del \dot{\phi}(\tau_0) }.
\ee

On the cotangent bundle of a supermanifold an odd canonical
two form is known to exist when
it has an equal number of odd and even coordinates
\cite{leites}. Thus we can define
an ``odd Poisson bracket'' (antibracket)
\be
\label{abrac}
(f,g) \equiv
\fr{\del_r f}{\del \phi}
\fr{\del_l g}{\del \Pi} -
\fr{\del_r f}{\del \Pi}
\fr{\del_l g}{\del \phi},
\ee
where $\del_r$ and $\del_l$ indicate the right and the left derivatives.
In this phase space odd time evolution is given by the Grassmann-even
hamiltonian $S$:
\be
\fr{\del f}{\del \tau_0}=(S,f).
\ee
Thus $S$ must satisfy
\be
\label{master}
\fr{\del S}{\del \tau_0} =(S,S) =0.
\ee
This is the master equation of Batalin and Vilkovisky.

Let us suppose that one can perform a Legendre
transformation to find the odd time hamiltonian $S$,
from the lagrangian $L$.
$L$ is taken to be  in first-order form (linear in $\dot{\phi}$),
and due to the fact that
\[
\Pi \equiv \phi^\star,
\]
where ``$^\star$" denotes the conjugation mapping fields into antifields
and vice versa, one can see that there are two candidates:
\begin{eqnarray}
i ) &     L= & \sig \dot{\phi}- H_{01} (\phi , \sig ), \lb{1st} \\
ii ) &  L=  & \phi \dot{\phi} - H_{02} (\phi )  .  \lb{2nd}
\end{eqnarray}
Multiplication of          fields is defined such that
the product possesses a definite
ghost number, e.g. ghost number of $L$ must be 0.
On the other hand when we deal with the components $\phi_i$,
\[
\phi_i \dot{\phi}_j \equiv  \phi_j \dot{\phi}_i .
\]

The case (\ref{2nd}) is  similar to the case (\ref{1st}):
in terms of the components $\phi_i$,
 $\phi\dot{\phi}$ which is defined to have ghost number 0,
 yields
 \[
 \phi_i\dot{\phi}_j,\     i\neq j.
 \]
 Then it is sufficent to deal only with the former case (\ref{1st}).
We will proceed  with odd time canonical formalism
following the analysis of Dirac for the constrained
hamiltonian systems\cite{dirac}.
In the case $i)$ there are two constraints
\[
\psi_1 \equiv \Pi_\phi - \sig =0,\   \psi_2 \equiv \Pi_\sig =0.
\]
The only non-vanishing antibracket of the constraints is
\[
(\psi_1 , \psi_2 )= -1.
\]
Define the extended hamiltonian as
\[
H_{e1}\equiv H_{01} + \psi_i \la_i ,
\]
so that $\dot{\psi}_i|_{\psi=0} =0$ will lead to
\[
\la_1 = -\fr{\del_l H_{01}}{\del \phi},\
\la_2 =  \fr{\del_l H_{01}}{\del \sig}.
\]
Therefore we can eliminate the constraints $\psi_i$,
and the odd time hamiltonian reads $H_{01}(\phi , \Pi_\phi )$.
Thus, we  conclude that as in
the normal canonical formulation  one can directly read
from the first order lagrangian, the non-vanishing
odd Poisson brackets and the related hamiltonian.

Now, we would like to give a method to find   the odd time lagrangian,
which is in the form discussed above.

To gather the original fields and the ghosts one can
extend differential forms to include also the ghost number. This
can be achieved by
generalizing the exterior derivative as \cite{cc}
\be
\lb{dtil}
d \rightarrow  \ti{d} \equiv d + \del / \del \tau_0 .
\ee
In order to utilize  this generalization of $d$,
as well as the odd time canonical formulation discussed
above, to find the minimal
solution of the master equation we should follow the following
procedure.

$i )$ If the started gauge theory is not already first order
in $d$ and the terms containing $d$ are not bilinear in fields,
one should find an equivalent formulation of it possessing these
properties.

$i i )$  Perform the change given in (\ref{dtil}) and generalize the
original fields to include also the ghosts and antifields which
possess the same order with them in terms of $\tilde{d}$.
(The ghost content of the theory should be found by analyzing the
related gauge invariance and the proper solution condition of
Batalin and Vilkovisky.)

$i i i )$ In terms of the Legendre transformation find the
related odd time  hamiltonian.

Of course, the crucial point    is to find the
conditions which should be satisfied such that the
odd time  hamiltonian
following from  this procedure
is a solution of the master equation.

To have a unified notation let us
deal with the case
given in (\ref{2nd}). Thus $(\phi , \phi )= 1$, and the master equation can
be written as
\[
(S,S)=\fr{\del S}{\del \phi_i}\fr{\del S}{\del \phi_j} g_{ij} =0,
\]
with an appropriate metric $g_{ij}$.

The form of the odd time hamiltonian
$H_0$, is the same with the original lagrangian $L_0$,
which is gauge invariant
\[
\fr{\del L_0}{\del \Phi_a} R^\al_a=0,
\]
where $\Phi_a$ indicate the original fields;
$\phi =\Phi + \cdots $, and $R_a^\al$ are the
gauge generators. Hence the odd time hamiltonian $H_0$,
will satisfy
\be
\fr{\del H_0}{\del \phi_i} \D \phi_i =0,
\ee
where
\[
\D  \phi_i = \ti{R}_i^j\phi_j.
\]
$\ti{R}$ is the generalization of $R$. If it is possible
to write
\be
\lb{cond}
\ti{R}_i^j\phi_j = g_{ij}\fr{\del H_0}{\del \phi_j},
\ee
one can conclude that the odd time hamiltonian is the minimal
solution of the master equation
\be
\lb{sol}
H_0 =S.
\ee
Of course, there may be some other theories whose odd time
hamiltonian satisfies (\ref{sol}), because of some other
conditions.

To clarify the procedure outlined above, let us  see some applications
of it.

\vspace{.2in}

{\bf 1) Yang-Mills Theory}

\vspace{.2in}

It is defined in terms of the second order action (we suppress
$Tr$)
\be
\lb{ym}
L_0= \int d^4x\   F_{\mu \nu}  F^{\mu \nu},
\ee
where
$F=d \wedge A + (1/2)A \wedge A$. The theory given by
\be
\lb{eym}
L=\int d^4x\   ( B_{\mu \nu}   F^{\mu \nu}
-\fr{1}{2} B_{\mu \nu} B^{\mu \nu} ),
\ee
is equivalent to  (\ref{ym}) on mass-shell, and moreover it
is first order
in $d$. (\ref{eym}) is invariant under the
infinitesimal gauge transformations
\[
\D A_\mu =D_\mu \La\ ,\  \D B_{\mu \nu}= [ B_{\mu \nu}, \La ],
\]
where $D=d + [A,\ ]$ is the covariant derivative. They are irreducible, so
that for the covariant quantization we
need only (in the minimal sector) the ghost field
$\eta$, which possesses ghost number 1.

Upon performing the change (\ref{dtil}) and generalizing
the fields of (\ref{eym}) as
\[
A  \rightarrow \ti{A}, \ B \rightarrow \ti{B},
\]
one obtains
\be
\ti{L}=\int d^4x\   [\ti{B} (d\ti{A} + \del \ti{A} /\del \tau_0
+\ti{A}\ti{A}) -\fr{1}{2} \ti{B}\ti{B} ],
\ee
which is defined to possess 0 ghost number.
Order of the extended forms $\ti{A}$ and $\ti{B}$, respectively,
are 1 and 2, and their first components
are $A_\mu$, $B_{\mu \nu}$. The definition of odd time canonical
momentum yields
\[
\Pi_{\ti{A}} =\ti{B},\  \Pi_{\ti{B}} =0.
\]
Thus odd time canonical hamiltonian is
\be
\lb{ochym}
H=\int d^4x\   [- \Pi_{\ti{A}} (d\ti{A}+ \ti{A}\ti{A})
+\fr{1}{2} \Pi_{\ti{A}}\Pi_{\ti{A}}].
\ee

By using the fact that
\[
N_{gh}(\phi) + N_{gh}(\phi^\star ) =-1,
\]
where $N_{gh}$ denotes the ghost number,
we write the generalized fields as
\[
\begin{array}{lcl}
\ti{A} & =  &  A_{(1+0)}+\eta_{(0+1)} + B_{(2-1)}^\star , \\
\Pi_{\ti{A}} & = & B_{(2+0)}+ A^\star_{(3-1)}+ \eta^\star_{(4-2)},
\end{array}
\]
where the first number in paranthesis is the order of d-forms
and the second is the ghost number. Here $``^\star"$ indicates
the antifields as well as the Hodge-map. Substitution of these
in (\ref{eym}) and using the property of the multiplication
that the product is different
from zero only when its ghost number vanishes, we
get
\be
\lb{syme}
H=-\int d^4x\ (B_{\mu \nu}F^{\mu \nu} +
B^{\mu\nu} [\eta , B_{\mu\nu}^\star ]
+A_\mu^\star D^\mu \eta + \eta^\star [\eta ,\eta ]
-\fr{1}{2} B_{\mu\nu} B^{\mu\nu}).
\ee
We may perform a partial gauge fixing $B^\star =0$, and then
use the equations of motion with respect to $B_{\mu\nu}$ to obtain
\[
H\rightarrow S=- \int d^4x\   (F_{\mu \nu}  F^{\mu \nu}
+A^\star_\mu    D^\mu \eta
+\eta^\star [\eta ,\eta] ),
\]
which is the minimal solution of the master equation for
Yang-Mills theory.

\vspace{.2in}

{\bf 2) Massive Abelian Theory in Stueckelberg Formalism }

\vspace{.2in}

It is defined in terms of the second order lagrangian
\be
\lb{mts}
L_{0}=\int d^4x\   [\fr{1}{2} F_{\mu \nu}
F^{\mu \nu} +m^2 (A_\mu -m^{-1}\del_\mu v)
(A^\mu -m^{-1}\del^\mu v)],
\ee
where $F=d \wv A$.
The action linear in $d$, and equivalent to (\ref{mts}) on mass-shell
is
\be
\lb{emts}
L=\int d^4x\   [\fr{1}{2} B_{\mu \nu} (d\wv A)^{\mu \nu}
-\fr{1}{4}B_{\mu \nu} B^{\mu \nu}
+m (A_\mu -m^{-1}\del_\mu v) K^\mu   -\fr{1}{2}K_\mu K^\mu ].
\ee
It is invariant under the gauge transformations
\[
\begin{array}{ll}
\D A_\mu = \del_\mu \La , &  \D B_{\mu \nu } =0, \\
\D v =m\La , &   \D K_\mu =0,
\end{array}
\]
which are irreducible, so that in the minimal sector
there is only one ghost: $\eta$.
By performing the change (\ref{dtil}) and replacing the
fields with their generalized ones,  we can see that
\be
\Pi_{\ti{A}} =\fr{1}{2}\ti{B},\
\Pi_{\ti{v}} =-\ti{K},\
\Pi_{\ti{B}} =0,\
\Pi_{\ti{K}} =0.
\ee
Following the general procedure outlined above we find
the odd time hamiltonian as
\be
H=\int d^4x\   [- \Pi_{\ti{A}} d\ti{A} + \Pi_{\ti{A}}\Pi_{\ti{A}}
+m(\ti{A} -m^{-1}d\ti{v})\Pi_{\ti{v}} +\fr{1}{2}\Pi_{\ti{v}}\Pi_{\ti{v}}].
\ee
The generalized fields are
\[
\begin{array}{lcl}
\ti{A} & =  &  A_{(1+0)}+\eta_{(0+1)} + B_{(2-1)}^\star , \\
\Pi_{\ti{A}} & = & B_{(2+0)}+ A^\star_{(3-1)}+ \eta^\star_{(4-2)}, \\
\ti{v} & =  & v_{(0+0)}+ K_{(1-1)}, \\
\Pi_{\ti{v}} & = & K_{(3+0)}+ v^\star_{(4-1)}.
\end{array}
\]
By respecting the rules of multiplication one finds in components
\[
H= -L -\int d^4x\   [ A^\star_\mu \del^\mu \eta  -m\eta v^\star ].
\]
We can eliminate $B$ and $K$ by using their equations of motion to
obtain
\[
H\rightarrow S=\int d^4x\   [
-\fr{1}{2} F^2_{\mu \nu} - m^2 (A_\mu -m^{-1}\del_\mu v)^2
- A_\mu^\star\del^\mu \eta  +m\eta v^\star ].
\]
Indeed, this is the minimal solution of the master equation
for the theory given by (\ref{mts}).

\pagebreak

{\bf 3) Relativistic Particle}

\vspace{.2in}

In terms of the canonical variables
satisfying the Poisson bracket relation $\{ p_\mu ,q^\nu \} =\D^\nu_\mu$,
relativistic particle is given by
\be
\lb{rp}
L_0 = -\int   (p \cdot dq -\fr{1}{2}e p \cdot p ),
\ee
where $dq^\mu = \del_t q^\mu  dt$.
A variable possesses two different grading: one of them is due
to one dimensional manifold of $t$ and the other one is
related to  space-time manifold.

(\ref{rp}) is invariant under
\[
\D q^\mu =p^\mu \La ,\   \D p =0 , \    \D e =\del_t \La .
\]

Now, we perform the change (\ref{dtil}) and generalize the fields
as
\[
q,\ e \rightarrow  \ti{q};\  p \rightarrow \ti{p}.
\]
$q$ and $e$ are treated on the same footing due to the fact
that there is not $de$ term in (\ref{rp}).
The first component
of $\Pi_{\ti{e}}$ would be vanishing, so that it behaves like
a component of a field, i.e. $\ti{q}$. Hence the odd
time lagrangian
is
\be
\lb{rpe}
L=-\int [\ti{p} d\ti{q} +\ti{p} \del\ti{q}/\del \tau_0
-\fr{1}{2}\ti{q}\ti{p}^2],
\ee
where
\[
\begin{array}{lcl}
\ti{q} & =  &  q^\mu_{(1+0+0)}+e_{(0+1+0)}  +\eta_{(0+0+1)}+
p^{\star \mu}_{(1+1-1)},  \\
\ti{p} & =  &  p_{\mu (d-1+0+0)} +q^\star_{\mu (d-1+1-1)}
+e^\star_{(d+0-1)} + \eta^\star_{(d+1-2)}.
\end{array}
\]
The numbers in the paranthesis
indicate, respectively, grading due to space-time,
grading due to 1-dimensinal manifold and ghost number.

By calculating the product in components one can show that the
odd time hamiltonian yields
\[
H=\int dt\  [p\cdot \del_t q +e^\star \del_t \eta -\fr{1}{2}
e p^2 + q^\star \cdot p \eta ],
\]
which is the minimal solution of the master equation
for the relativistic particle.

\pagebreak

{\bf 4) The Self-interacting Antisymmetric Tensor Field}

\vspace{.2in}

The action\cite{ft} (we suppress $Tr$)
\be
\lb{si}
L_0 =\int d^4x\   [B_{\mu \nu}(d\wv A+\fr{1}{2}A\wv A)^{\mu \nu}
- \fr{1}{2} A_\mu A^\mu ],
\ee
is invariant under the transformations
\[
\D B_{\mu \nu } =\ep_{\mu \nu \rho \sigma }D^\rho \La^\sigma ,\
\D A_\mu =0 ,
\]
is analysed in terms of the BRST methods in Ref.\cite{af}.
If we set $\La_\mu = D_\mu \al$, the gauge transformation
vanishes on shell $\D B|_{F=0}=0$. This is a first-stage
reducible theory, hence we need to introduce the ghost fields
\[
C_0^\mu , \ C_1 ;\  N_{gh}(C_0^\mu )=1 , \ N_{gh}(C_1)=2.
\]
By following the general
procedure we find the odd time lagrangian
\be
\lb{afe}
L = \int d^4x\   [ \ti{B} (d\ti{A}  +\del \ti{A}/ \del \tau_0 +
\ti{A}\ti{A})  -\fr{1}{2} \ti{A}\ti{A}],
\ee
where the generalized fields are
\[
\begin{array}{ll}
\ti{A} = &  A_{(1+0)} +B^\star_{(2-1)} +C_{0(3-2)}^\star
+C_{1(4-3)}^\star ,\\
\ti{B} = &  B_{(2+0)} +A^\star_{(3-1)} +C_{0(1+1)} +C_{1(0+2)}.
\end{array}
\]
In terms of the components one can see that
the odd time hamiltonian is
\be
\lb{ath}
H=-\int d^4x\ \{ B_{\mu\nu}F^{\mu\nu} +
\ep_{\mu \nu \rho \sig }C_0^\mu D^\nu B^{\star \rho \sig}
+C_1 D^\mu C^\star_{0\mu} +\fr{1}{2}\ep^{\mu \nu \rho \sig }
C_1 [B^\star_{\mu \nu }, B_{\rho \sig}^\star ] -\fr{1}{2} A_\mu A^\mu \} .
\ee
This is the minimal solution of the
master equation of the theory defined by
(\ref{si})\cite{af}.

\vspace{.3in}

\begin{center}
Acknowledgments
\end{center}

This work was partly supported by Turkish Scientific and Technological
Research Council (TUBITAK).

\pagebreak

\end{document}